# OPTION PRICING UNDER NORMAL DYNAMICS WITH STOCHASTIC VOLATILITY

Matta Uma Maheswara Reddy†

† Department of Economics, Peking University HSBC Business School, Shenzhen, China; matta.umamaheswarareddy@gmail.com

**ABSTRACT**

In this paper, we derive the price of a European call option of an asset following a normal process assuming stochastic volatility. The volatility is assumed to follow the Cox–Ingersoll–Ross (CIR) process. We then use the fast Fourier transform (FFT) to evaluate the option price given we know the characteristic function of the return analytically. We compare the results of fast Fourier transform with the Monte-Carlo simulation results of our process. Further, we present a numerical example to understand the normal implied volatility of the model.

# 1. Introduction

The Black-Scholes (Black, 1973) model for option pricing is criticised often for its failure in explaining the case of varying implied volatilities (the volatility smile). Previous literature (Hull & White, 1987; Stein & Stein, 1991) addressed the issue by assuming stochastic volatility. Heston (1993) presented a semi-closed form solution for the Black-Scholes model under stochastic volatility with the volatility following the Cox-Ingersoll-Ross model. Given the Black-Scholes model is based on geometric-Brownian motion, the above studies focus on such models. Observing fatter tails in stock distributions and the fact that stock prices cannot be negative, tend to push us into adopting the lognormal distribution for stock prices, and hence, the call prices. The Black-Scholes model explains the movement of stocks and their derivatives well. We see, that on the other side, other financial asset classes cannot be explained well using the Black-Scholes model, given the limitations.

The Normal model overcomes some of these limitations and can help us analyse the asset prices better compared to the Black-Scholes model. Especially in the aspect of measuring volatility, the Black-Scholes model measures the relative change in the asset value whereas the Normal model measures the absolute change in the asset value. Studying the normal model helps us understand the price changes and would be useful for hedging purposes given the hedging ratio and delta being different for the Normal model and the Black-Scholes model. Though the Normal model is believed to be one of the first option-pricing models, the number of studies on stochastic volatility models based on arithmetic Brownian motion (normal stochastic volatility models) is scant.

Bachelier (1900) analysed the arithmetic Brownian motion for the first time in his doctoral thesis in mathematics. It has been observed that the arithmetic Brownian motion explains some assets such as interest rate (Levin, 2004), inflation rate (Kenyon, 2008) and spread options better compared to the geometric Brownian motion. For some recent studies on the normal model, please refer to Brooks & Brooks (2017) and Schachermayer & Teichmann (2008). Despite the normal stochastic volatility models having a vital background, we see there are rarely any studies to address it. This paper aims at solving the stochastic volatility model followed by an evaluation of the European call option price in this process using fast Fourier transform (FFT).

There are several research questions we try to answer in this work. It starts with solving the normal model under stochastic volatility. Once we find a solution to the equation, we use the fast Fourier transform to evaluate the option price. Our contribution is to consider a normal stochastic volatility model and applying the fast Fourier transform to evaluate the option price. We then compare the results of the FFT with the Monte Carlo simulation results and compare the accuracy. In the numerical analysis part, we discuss the implied volatility of the model.

## 2. Normal Stochastic Volatility Model

Let $x(t)$ be the spot price of an asset at time $t$. It follows the diffusion equation:

$$dx(t) = rdt + \sqrt{v(t)}dz_1(t) \tag{2.1}$$

Let the volatility of the process be stochastic following the equation:

$$dv = (a - bv)dt + \sigma\sqrt{v(t)}dz_2(t) \tag{2.2}$$

where $z_1(t), z_2(t)$ are Brownian motions under the forward measure with correlation $\rho$. $a/b$ is the long variance; the value of $v(t)$ tends to $a/b$ as $t \to \infty$. $b$ is the mean reversion rate; the rate at which $v(t)$ reverts to $a/b$.

Consider any twice-differentiable function $f(x, v, t)$ that is the conditional expectation of some function of $x$ and $v$ at a later date $(T)$, $g(x(T), v(T))$:

$$f(x, v, t) = E[g(x(T), v(T)) | x(t) = x, v(t) = v] \tag{2.3}$$

Ito's lemma shows that

$$df = \frac{\partial f}{\partial t}dt + \frac{\partial f}{\partial x}dx + \frac{\partial f}{\partial v}dv + \frac{1}{2}v\frac{\partial^2 f}{\partial x^2}dt + \frac{1}{2}\sigma^2 v\frac{\partial^2 f}{\partial v^2}dt + \frac{1}{2}\rho\sigma v\frac{\partial^2 f}{\partial v \partial x}dt \tag{2.4}$$

$$\equiv \left(\frac{\partial f}{\partial t} + r\frac{\partial f}{\partial x} + (a - bv)\frac{\partial f}{\partial v} + \frac{1}{2}v\frac{\partial^2 f}{\partial x^2} + \frac{1}{2}\sigma^2 v\frac{\partial^2 f}{\partial v^2} + \frac{1}{2}\rho\sigma v\frac{\partial^2 f}{\partial v \partial x}\right)dt$$

$$+ \sqrt{v}\frac{\partial f}{\partial x}dz_1 + \sigma\sqrt{v}\frac{\partial f}{\partial v}dz_2$$

Using iterated expectations, we see that $f$ should be a martingale

$$E[df] = 0 \tag{2.5}$$

Applying (3.5) to (3.4)

$$\frac{\partial f}{\partial t} + r\frac{\partial f}{\partial x} + (a - bv)\frac{\partial f}{\partial v} + \frac{1}{2}v\frac{\partial^2 f}{\partial x^2} + \frac{1}{2}\sigma^2 v\frac{\partial^2 f}{\partial v^2} + \frac{1}{2}\rho\sigma v\frac{\partial^2 f}{\partial v \partial x} = 0 \tag{2.6}$$

The terminal condition from (3.3)

$$f(x, v, T) = g(x, v) \tag{2.7}$$

This is where we take $g(x, v) = e^{i\alpha x}$, hence by associating the function $g(x, v)$ as the characteristic function of $x$. Characteristic function is defined as the Fourier transform of the probability destiny of a random variable. Characteristic functions are extremely handy in scenarios when the density function is unknown. In such cases, one can analytically determine the density function by inverse Fourier transforming the characteristic function.

Solving for the characteristic function, we start by "guessing" the functional form

$$f(x, v, t) = exp[C(T - t) + D(t - t)v + i\omega x] \qquad (2.8)$$

Substituting (3.8) in (3.6)

$$f\left[-\frac{\partial C}{\partial t} - v\frac{\partial D}{\partial t} + \frac{1}{2}v(i\omega)^2 + \rho\sigma vD(i\omega) + \frac{1}{2}\sigma^2 vD^2 + ri\omega + (a - bv)D\right] = 0 \qquad (2.9)$$

Since $f$ is a non-zero function, the argument of the function should be zero. We have

$$\left[-\frac{\partial C}{\partial t} + aD + ri\omega\right] + v\left[-\frac{\partial D}{\partial t} + \frac{1}{2}(i\omega)^2 + \rho\sigma D(i\omega) + \frac{1}{2}\sigma^2 D^2 - bD\right] = 0 \qquad (2.10)$$

Without loss of generality, assuming non-zero volatility, we have the following differential equations:

$$\begin{aligned}-\frac{1}{2}\omega^2 + \rho\sigma\omega iD + \frac{1}{2}\sigma^2 D^2 - bD - \frac{\partial D}{\partial t} &= 0 \\ r\omega i + aD - \frac{\partial C}{\partial t} &= 0\end{aligned} \qquad (2.11)$$

We have the solution (Refer Appendix for the derivation)

$$C(\tau;\omega) = -ri\omega\tau + \frac{a}{\sigma^2}\left[(b - \rho\sigma i\omega - d)\tau - 2ln\left(\frac{1 - ge^{-d\tau}}{1 - g}\right)\right]$$

$$D(\tau;\omega) = \frac{b - \rho\sigma i\omega - d}{\sigma^2}\left(\frac{1 - e^{-d\tau}}{1 - ge^{-d\tau}}\right)$$

where

$$g = \frac{b - \rho\sigma i\omega - d}{b - \rho\sigma i\omega + d}; d = \sqrt{(b - \rho\sigma i\omega)^2 + \sigma^2\omega^2} \qquad (2.12)$$

# 3. Fourier transform of an option price (and FFT computation)

We follow the method followed by Carr-Madan[10] to evaluate the call option price for the Normal Stochastic Volatility model using fast Fourier transform. Let $C_T(K)$ denote the call price of an option with strike $K$ and maturity $T$. Let $q_t(S)$ be the probability density function of the spot price $S_T$. The characteristic function of this distribution is

$$\phi_T(u) = \int_{-\infty}^{\infty} e^{iuS} q_T(S) dS \qquad (3.1)$$

The relation between the initial call value $C_T(K)$ and the probability density function $q_T(S)$

$$C_T(K) = \int_K^\infty e^{-rT}(S-K)q_T(S)dS \qquad (3.2)$$

The function $C_T(K)$ would not be square-integrable as $K$ tends to $-\infty$. The function is therefore dampened to obtain a square-integrable function. The modified call option price $c_t(K)$ is defined as

$$c_T(K) = exp(\alpha K)C_T(K) \qquad (3.3)$$

where $\alpha > 0$.

The Fourier transform of $c_T(K)$ is defined by

$$\psi_T(v) = \int_{-\infty}^\infty e^{ivK}c_T(K)dK \qquad (3.4)$$

We develop an expression for $\psi_T(v)$ using $\phi_T$ and obtained the call prices using the inverse_Fourier transform

$$C_T(K) = \frac{exp(-\alpha K)}{2\pi}\int_{-\infty}^\infty e^{-ivK}\psi_T(v)dv = \frac{exp(-\alpha K)}{\pi}\int_0^\infty e^{-ivK}\psi_T(v)dv \qquad (3.5)$$

We get the second equability from the fact the that expression $e^{ivK}c_T(K)$ is an odd function in its imaginary part given that $c_T(K)$ is real. We determine the expression for $\psi_T(v)$ as follows

$$\psi_T(v) = \int_{-\infty}^\infty e^{ivK}c_t(K)dK = \int_{-\infty}^\infty e^{ivK}e^{\alpha K}C_T(K)dK = \int_{-\infty}^\infty e^{ivK}e^{\alpha K}\int_K^\infty e^{-rT}(S-K)q_T(S)dSdK$$

$$\equiv \int_{-\infty}^\infty e^{-rT}q_T(S)\int_{-\infty}^S (S-K)e^{\alpha K}e^{ivK}dKdS = \int_{-\infty}^\infty e^{-rT}q_T(S)\int_{-\infty}^S (S-K)e^{(\alpha+iv)K}dKdS$$

Taking $(\alpha + iv) = a$, we solve the equation by integrating in parts

$$\equiv \int_{-\infty}^\infty e^{-rT}q_T(S)\frac{e^{(\alpha+iv)S}}{(\alpha+iv)^2}dS = \frac{e^{-rT}}{(\alpha+iv)}\left[\phi_T(-i(\alpha+iv))\right]$$

$$= \frac{e^{-rT}}{(\alpha+iv)}\left[\phi_T(v-i\alpha)\right] \qquad (3.6)$$

Fast Fourier transform (FFT) is a computation algorithm that helps in computing the following sum efficiently

$$w(k) = \sum_{j=1}^N e^{-i\frac{2\pi}{N}(j-1)(k-1)}x(j) \text{ for } k = 1,2,3,...,N \qquad (3.7)$$

From the above expression for the call price

$$C_T(K) = \frac{exp(-\alpha K)}{\pi} \int_0^\infty e^{-ivK} \psi_T(v) dv \qquad (3.8)$$

We approximate the call price using the trapezoidal rule and set $v_j = \eta(j-1)$

$$C_T(K) \approx \frac{exp(-\alpha K)}{\pi} \sum_{j=1}^N e^{-iv_j K} \psi_T(v_j) \eta \qquad (3.9)$$

The upper limit for the integral would be $L_u = N\eta$

Let us take the strike price we are interested in be $K_0$. The FFT returns $N$ values of $K$, with the mean $K_0$. Taking the spacing to be $\lambda$, the values of $K$ are

$$K_u = -b + \lambda(u-1) \text{ for } u = 1,2,3,...,N \qquad (3.10)$$

Equation (3.10) gives us strike levels in the range of $(-b + K_0)$ to $(b + K_0)$ where $b = \frac{1}{2} N\lambda$

Substituting (3.9) into (3.10), we get

$$C_T(K) \approx \frac{exp(-\alpha K)}{\pi} \sum_{j=1}^N e^{-iv_j[-b+\lambda(u-1)]} \psi_T(v_j) \eta \qquad (3.11)$$

As $v_j = \eta(j-1)$, we have

$$C_T(K) \approx \frac{exp(-\alpha K)}{\pi} \sum_{j=1}^N e^{-i\lambda\eta(j-1)(u-1)} e^{ibv_j} \psi_T(v_j) \eta \qquad (3.12)$$

From (3.7), to apply FFT, we need $\lambda\eta = \frac{2\pi}{N}$. That would imply that if we choose a small $\eta$ for obtaining a good integration grid, we would have call prices between strikes with large spacings. Therefore, in order to obtain call prices for various strikes with little spacing, we choose large values of $\eta$. Using Simpson's rule to evaluate the (3.12), we get

$$C_T(K) = \frac{exp(-\alpha K)}{\pi} \sum_{j=1}^N e^{-i\lambda\eta(j-1)(u-1)} e^{ibv_j} \psi_T(v_j) \frac{\eta}{3}[3 + (-1)^j - \delta_{j-1}] \qquad (3.13)$$

where $\delta_n$ is the Kronecker delta function. To compute the call prices, one has to make appropriate choices for $\eta$ and $\alpha$. The summation in (3.13) is the exact application of the fast Fourier transform.

# 4. Fast Fourier transform v/s Monte-Carlo Simulation

In this section, the FFT approximation of the call prices is compared with the Monte-Carlo simulation of the model. We run the simulations for fast Fourier transform for various values of *N*. The code for FFT is mentioned in the appendix. The FFT and Monte-Carlo simulations are performed on a 2017 MacBook Pro with Intel Core i5 processor (2.3GHz dual-core) and an 8GB RAM. The results of the FFT approximation are compared with the results of Sidani[11].

**Table 1**: FFT solution for the closed-form solution vs. Monte Carlo (MC). The option specs are: $S_0$= -10 bps; $\tau = 1$. The model parameters used: $a = 0.5S_0^2; b = 1; \sigma = 0.25; \rho = -0.09; v_0 = 0.09$. We take $\eta = 1$ and calculate the call option price over a range of $\alpha$'s with $\alpha_0 = 5$ with an increment of 0.1 for 500 points (we have $\alpha_{500} = 55$) and take the mean of those values.

| Strike/N | 1024 | 2048 | 4096 | 8192 | 16384 | 32768 | MC |
|---|---|---|---|---|---|---|---|
| -0.0005 | 0.08870 | 0.09046 | 0.09135 | 0.09179 | 0.09201 | 0.09213 | 0.09220 |
| 0 | 0.08842 | 0.09018 | 0.09106 | 0.09150 | 0.09172 | 0.09184 | 0.09197 |
| 0.0005 | 0.08813 | 0.08988 | 0.09077 | 0.09121 | 0.09144 | 0.09155 | 0.09152 |

In the table above (Table 1), we observe accurate results with the increment in the step count for the fast Fourier transform. In the table below (Table 2), we compare the computation times for FFT and the MC simulation.

**Table 2**: Running time comparison - FFT vs Monte-Carlo. 'N' denotes the number of steps in FFT; 'ts' denotes the number of time-steps in the Monte-Carlo simulation (with 20 repetitions). The prices are calculated for 20 strike values.

|  | FFT | Monte-Carlo |
|---|---|---|
| N = 2048, ts = 10000 | 0.003 | 2.725 |
| N = 4096, ts = 20000 | 0.004 | 6.519 |
| N = 8192, ts = 30000 | 0.007 | 10.708 |
| N = 16384, ts = 40000 | 0.012 | 15.545 |
| N = 32768, ts = 50000 | 0.018 | 20.466 |

It is observed that the fast Fourier transform provides an accurate price taking much less computation time. This stands in line with the general observation for other option pricing models comparing the results of FFT and Monte-Carlo simulation.

# 5. Implied Volatility

A numerical example on the implied volatility of the Normal Stochastic Volatility model is presented in this section.

**Figure 1**: Normal Implied Volatility surface for the model with the parameters: $\rho = -0.9; \sigma = 0.25; \phi = 0.25; a = 5*10^{-7}; b = 1; v_0 = 0.09; S_0 = 1; K = 0.7 - 1.9; T = 0.6 - 2$ years

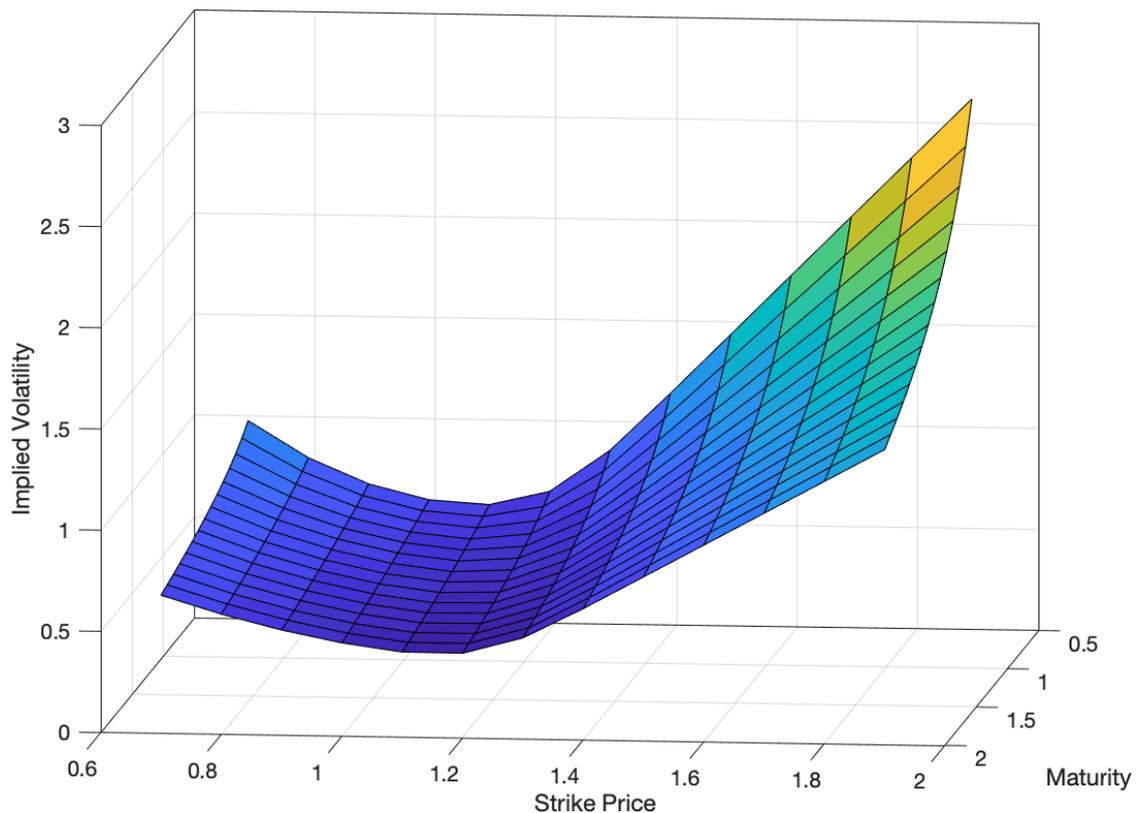

It is observed that the implied volatility goes above 1 (i.e. above 100%). These values can be associated with a higher volatility-of-volatility value and can therefore come down when $\sigma$ is decreased. This characteristic property of the model is highly resourceful when explaining the out of the money options (OTM). Considering the out of the money options, given the probability of price jumping to these extremes, the Black-Scholes implied volatilities fail to account the high volatility in the markets that bring up such changes. The Normal implied volatility accounts for such changes better than the Black-Scholes implied volatility. This is one of the important application of the Normal model to real markets.

In Figure 2, we study the change in shape of implied volatility surface with respect to the correlation factor. The minima of the curve tends towards the left as the correlation factor increases from -0.9 to 0.9.

**Figure 2**: Normal Implied Volatility surface for the model with the parameters: $\sigma = 0.25; \phi = 0.25; a = 5 * 10^{-7}; b = 1; v_0 = 0.09; S_0 = 1; K = 0.7 - 1.9; T = 0.6 - 2; \rho = [-0.9, 0, 0.9]$ from top to bottom respectively

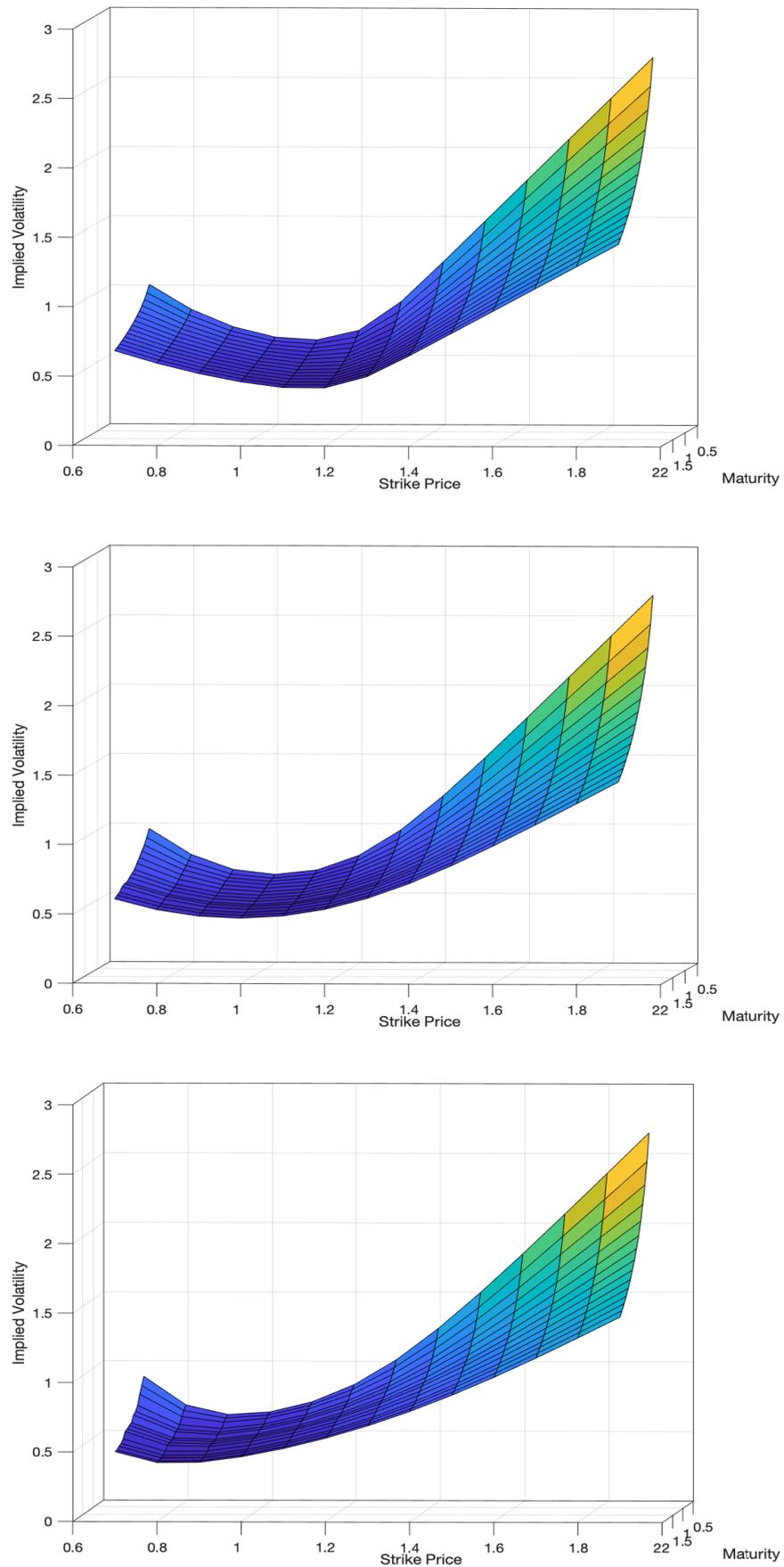

# 6. Conclusion

In this work, a solution for a European call option following the arithmetic Brownian motion under stochastic volatility is derived using Fourier transformation. We use the fast-Fourier transform (FFT) to analytically determine the value of the option price under the given dynamics. It is observed that the output of FFT comes very close to the Monte-Carlo simulation results of the system. This is followed by studying the implied volatility surface of the model which could potentially explain extreme changes in the market better than the Black-Scholes implied volatility.

# Appendix

**Lemma**: The solution to the systems of ODE's specified by (2.11) is given by

$$C(\tau;\omega) = -ri\omega\tau + \frac{a}{\sigma^2}\left[(b - \rho\sigma i\omega - d)\tau - 2\ln\left(\frac{1 - ge^{-d\tau}}{1 - g}\right)\right]$$

$$D(\tau;\omega) = \frac{b - \rho\sigma i\omega - d}{\sigma^2}\left(\frac{1 - e^{-d\tau}}{1 - ge^{-d\tau}}\right)$$

where

$$g = \frac{b - \rho\sigma i\omega - d}{b - \rho\sigma i\omega + d}; d = \sqrt{(b - \rho\sigma i\omega)^2 + \sigma^2\omega^2}$$

**Proof**: We first solve for $D(\tau;\omega)$

$$-\frac{1}{2}\omega^2 + \rho\sigma\omega i D + \frac{1}{2}\sigma^2 D^2 - bD - \frac{\partial D}{\partial t} = 0$$

This changes to

$$-\frac{1}{2}\omega^2 + \rho\sigma\omega i D + \frac{1}{2}\sigma^2 D^2 - bD + \frac{\partial D}{\partial \tau} = 0 \text{ as } \tau = T - t$$

$$\frac{\partial D}{\partial \tau} = \frac{1}{2}\omega^2 - \rho\sigma\omega i D - \frac{1}{2}\sigma^2 D^2 + bD$$

$$\frac{\partial D}{\partial \tau} = -(-n + mD + lD^2)$$

where

$$l = \frac{1}{2}\sigma^2, m = b - \rho\sigma\omega i, n = \frac{1}{2}\omega^2$$

$$-\frac{\partial D}{\partial \tau} = l(D - q_+)(D - q_-)$$

with

$$q_\pm = \frac{-m \pm d}{2l} \text{ where } d = \sqrt{m^2 + 4ln}$$

Separating the variables on both sides and solving it as an ODE

$$\frac{dD}{l(D - q_+)(D - q_-)} = -d\tau$$

which can be written as

$$\left[\frac{1/(q_+ - q_-)}{l(D - q_+)} - \frac{1/(q_+ - q_-)}{l(D - q_-)}\right] dD = -d\tau$$

Integrating on both sides

$$\frac{\log(D - q_+)}{l(q_+ - q_-)} - \frac{\log(D - q_-)}{l(q_+ - q_-)} = -\tau + c_0$$

Since $l(q_+ - q_-) = d$ and we have $D(0; \omega) = 0$, we have

$$\frac{\log(-q_+)}{d} - \frac{\log(-q_-)}{d} = c_0$$

and hence

$$e^{(c_0 d)} = \frac{q_+}{q_-} \equiv g$$

Solving for $D$, we have

$$D(\tau; \omega) = q_+ \left(\frac{1 - e^{-d\tau}}{1 - ge^{-d\tau}}\right)$$

$$\equiv D(\tau; \omega) = \frac{-(b - \rho\sigma i\omega - d)}{\sigma^2}\left(\frac{1 - e^{-d\tau}}{1 - ge^{-d\tau}}\right)$$

We now solve for $C(\tau; \omega)$

$$r\omega i + aD - \frac{\partial C}{\partial t} = 0$$

This changes to

$$\frac{\partial C}{\partial \tau} = -(r\omega i + aD) \text{ where } \tau = T - t$$

Solving this as an ODE, we have

$$C(\tau; \omega) = -\int (r\omega i + aD) d\tau$$

$$= -r\omega i\tau - a\left[q_+\tau + \frac{2}{\sigma^2} ln\left(1 - ge^{-d\tau}\right)\right] + c_1$$

As $C(0; \omega) = 0$, we have

$$c_1 = \frac{2a}{\sigma^2} ln\left(1 - ge^{-d\tau}\right)$$

We finally have

$$C(\tau; \omega) = -ri\omega\tau + \frac{a}{\sigma^2}\left((b - \rho\sigma i\omega - d)\tau - 2\ln\left[\frac{1 - ge^{-d\tau}}{1 - g}\right]\right)$$